\documentclass[aps,prd,twocolumn,longbibliography,nofootinbib,showkeys]{revtex4-2}
\usepackage[left=2.5cm,right=2.5cm,top=2.4cm,bottom=2.4cm]{geometry}

\usepackage[english]{babel}
\usepackage[latin1]{inputenc}
\usepackage{enumerate}
\usepackage{amsmath}
\usepackage{amsthm}
\usepackage{amssymb}
\usepackage{graphicx}
\usepackage{floatflt}
\usepackage{fancybox}
\usepackage{appendix}
\usepackage{enumitem}
\usepackage{subfigure}
\usepackage{array}

\usepackage{hyperref}
\usepackage[b]{esvect}

\usepackage{color}

\renewcommand{\Re}{\textrm{Re}}

\newcommand{\om}{\omega}

\newcommand{\Gam}{\Gamma}

\newcommand{\id}{\mathbf{1}} 

\newcommand{\be} {\begin{equation}}
\newcommand{\ee} {\end{equation}}
\newcommand{\bsub}{\begin{subequations}}
\newcommand{\esub}{\end{subequations}}
\newcommand{\bea}{\begin{eqnarray}}
\newcommand{\eea}{\end{eqnarray}}
\newcommand{\bi} {\begin{itemize}}
\newcommand{\ei} {\end{itemize}}
\newcommand{\ben} {\begin{enumerate}}
\newcommand{\een} {\end{enumerate}}
\newcommand{\bmat} {\begin{pmatrix}}
\newcommand{\emat} {\end{pmatrix}}
\newcommand{\bal} {\begin{aligned}}
\newcommand{\eal} {\end{aligned}}
\newcommand{\btab}{\begin{tabular}}
\newcommand{\etab}{\end{tabular}}

\begin{document}
\selectlanguage{english}

\title{Single $\pi$-flux hosting topological defect modes in bilayer acoustic metamaterials}

\author{Renaud Cote} 
\email{cote@lma.cnrs-mrs.fr}
\affiliation{Aix Marseille Univ., CNRS, Centrale Marseille, LMA UMR 7031, Marseille, France}

\author{Marc Pachebat} 
\email{pachebat@lma.cnrs-mrs.fr}
\affiliation{Aix Marseille Univ., CNRS, Centrale Marseille, LMA UMR 7031, Marseille, France}

\author{Antonin Coutant}
\email{coutant@lma.cnrs-mrs.fr}
\affiliation{Aix Marseille Univ., CNRS, Centrale Marseille, LMA UMR 7031, Marseille, France}

\date{\today}

\begin{abstract}
	The bulk-boundary correspondence, which relates topological properties of a material in the bulk to the presence of robust modes localized on the edge, is at the core of the now mature field of topological wave physics. More recently, it was realized that in crystalline structures, certain types of defects can host localized modes, in which case the bulk-boundary correspondence has to be replaced by a bulk-defect correspondence. These defect-localized modes are expected to have robust properties owing to their topological origin. In this work, we show how to obtain topological defect modes in a lattice possessing both mirror and chiral symmetry. The defect is obtained by endowing a plaquette with a non-trivial gauge flux. We show that the bulk-defect correspondence is satisfied by introducing appropriate topological invariants. Moreover, the topological defect modes are shown to be highly robust to the introduction of symmetry-preserving disorder. The model is then realized in an acoustic system made of a bilayer network of tubes, and the presence of topological defect modes is experimentally clearly demonstrated. 
\end{abstract}


\maketitle


%
%

\section{Introduction}

The interplay between symmetries and topology has opened a vast domain of research, first in the context of electronic states in solid state physics~\cite{Hasan2010} and then in their counterpart for classical waves, from electromagnetic~\cite{Ozawa2019,Price2022} to acoustics or elastic waves~\cite{Ma2019,ZangenehNejad2020a}. The hallmark of topological wave systems is the presence of modes localized on the boundary, and that possess exceptional robustness against the presence of defects in the system. More recently, a new class of topological modes was exhibited: modes that are attached to a topological defect in an otherwise pristine lattice~\cite{Zhang2023}. A paradigmatic example is that of Majorana modes bounded to a vortex in a topological superconductor~\cite{Alicea2012}, which finds it origin in the Jackiw-Rossi mechanism~\cite{Jackiw1981}. This concept was later transposed to the realm of classical waves by designing a smooth vortex-like modulation of a honeycomb phononic crystal~\cite{Gao2019,Chen2019a,Gao2020}. 

In this work, we report the observation of a new class of topological defect modes in a two-dimensional lattice displaying both chiral and mirror symmetry. These topological modes arise in pairs in lattices with an isolated $\pi$-flux defect. A $\pi$-flux is obtained when some hopping coefficients of a tight binding model have negative signs, such that the accumulated phase around a plaquette is $\pi$. This corresponds to a so-called ``gauge flux''~\cite{Chen2023}. Lattices with $\pi$-fluxes have already been successfully used in the context of metamaterials to realize exotic wave systems such as higher order topological insulators~\cite{Benalcazar2017a,SerraGarcia2018,Mittal2019,Qi2020,Zhu2020}, or lattices displaying projective symmetries~\cite{Xue2022,Li2022,Chen2022,Chen2023,Li2023,Jiang2024}. However, isolated $\pi$-flux, and their ability to host defect localized topologically protected modes have received much less attention~\cite{Juricic2012,Kariyado2019}. 

Due to the presence of both chiral and mirror symmetries, the considered lattices have a topology characterized by a pair of topological invariants called mirror winding numbers~\cite{Kariyado2017,Coutant2024}. We show that isolated $\pi$-flux can host localized modes depending on the topological phase of the system. Surprisingly~\cite{Kariyado2019}, the topological defect modes arise in the ``trivial phase'' (vanishing invariants). This is explained by appealing to the bulk-defect correspondence using appropriately defined topological indices for the defect itself. 

The acoustic realization of the model is obtained by using a network of slender tubes, which have been shown to provide broadband and easy-to-control implementations of tight binding models~\cite{Zheng2019,Zheng2020,Coutant2020a,Coutant2021,Coutant2021a,Coutant2024}. The $\pi$-flux defect is then implemented by using a double layer network, where inter-layer couplings are used to obtain negative hopping coefficients. The presence the topological defect modes is demonstrated experimentally by comparing the frequency response function in a pristine lattice and in a lattice containing a $\pi$-flux defect.

%
%

\section{Topological defects}

\begin{figure}[htp]
	\centering
	\includegraphics[width=\columnwidth]{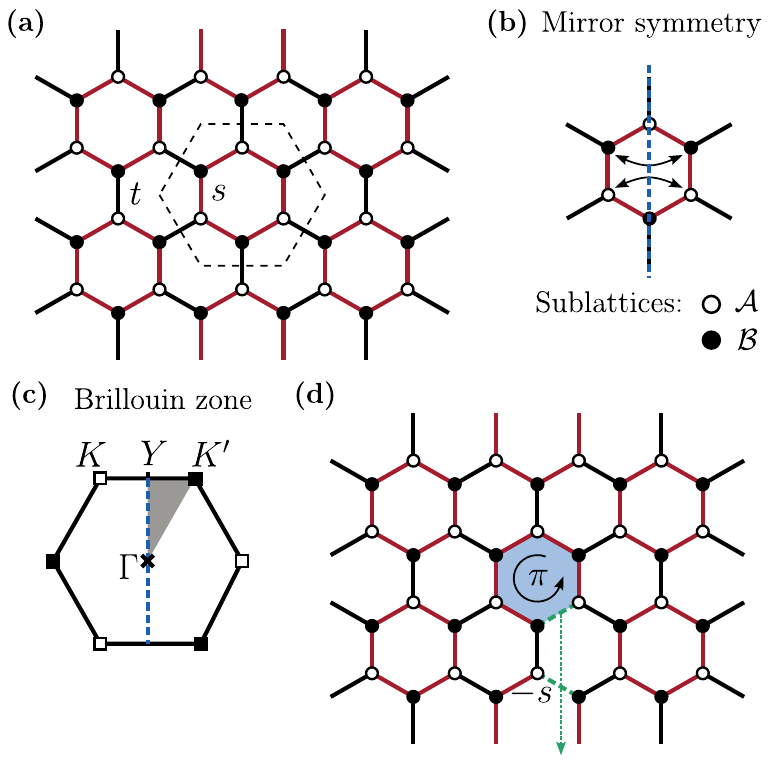} 
	\caption{(a) The Kekulé lattice model. The chosen unit cell is shown in dashed line. Black and white dots indicate the two sublattices $\mathcal A$ and $\mathcal B$. (b) symmetries of the model illustrated in a unit cell. (c) First Brillouin zone with high symmetry points ($K$, $K'$ and $Y$) and reduced Brillouin zone (in grey). The symmetry line is emphasized by a blue dashed line.} (d) A defect made of a hexagon threaded by a $\pi$-flux (filled with blue). In this work, the $\pi$-flux is obtained by flipping the signs of the hopping coefficients along the dashed line (negative hoppings are emphasized in green dashed).
	\label{Fig:Kekule_PiF} 
\end{figure}

The key ingredients of this work are the simultaneous presence of chiral and mirror symmetry that commute with each other. For the sake of the demonstration, we consider the Kekulé lattice model shown in Fig.~\ref{Fig:Kekule_PiF}-(a), and the symmetries illustrated in Fig.~\ref{Fig:Kekule_PiF}-(b). To confirm the generality of our findings, we show in Appendix~\ref{App:SquareLattice} how to obtain the same type of topological defect modes in a square lattice. The Kekulé model is described by the tight binding Hamiltonian 
\begin{equation} \label{eq:H_def}
	H = \sum_{\langle i,j\rangle} t_{ij} a_i^\dagger a_j, 
\end{equation}
where $\langle i,j\rangle$ denotes nearest neighbor on a honeycomb lattice, and $a_j$ is the annihilation operator on site $j$. The hopping coefficients $t_{ij}$ display a Kekulé modulation: intra-hexagon coefficients $t_{ij} = s$ differ from inter-hexagon coefficients $t_{ij} = t$, as illustrated in Fig.~\ref{Fig:Kekule_PiF}-(a). We also define $E$ the energy eigenvalue of $H$. 

This model and its edge modes have been thoroughly studied~\cite{Chamon2000,Liu2017a,Liu2019a,Bunney2022}, most often as an analogue of the quantum spin Hall effect~\cite{Wu2015,Wu2016,Liu2017b}, therefore relying on the hexagonal symmetry group $C6v$. In this work however, we use the combination of sublattice symmetry and mirror symmetry, which lead to a different type of topology, much less studied in the literature~\cite{Kariyado2017,Kariyado2019,Coutant2024}. Sublattice symmetry, also called chiral symmetry, is characterized by the chiral operator $\Gam = \mathrm{diag}(- \id_{\mathcal A}, \id_{\mathcal B})$, where $\mathcal A$ and $\mathcal B$ are the two sublattices. Chiral symmetry means that $\Gamma$ anti-commutes with the Hamiltonian $\Gam H + H \Gam = 0$. In addition, the mirror symmetry operation $M$ commutes with the Hamiltonian $M H - H M = 0$. Finally, we require the two symmetries to commute $\Gam M - M \Gam = 0$. This means that two mirror symmetric points must belong to the same sublattice. This imposes to choose $M$ as a mirror symmetry with an axis passing through summits of a hexagon (see Fig.~\ref{Fig:Kekule_PiF}-(b)). 

When this combination of symmetry is preserved, the system can display non-trivial topology characterized by mirror winding numbers~\cite{Kariyado2017,Coutant2024}. In short, the Bloch Hamiltonian commutes with $M$ along the high symmetry line $\Gamma-Y$ (see Fig.~\ref{Fig:Kekule_PiF}-(c)), and hence, splits onto symmetric and anti-symmetric sectors. This gives two effective 1D Hamiltonians, each of them with sublattice symmetry, and hence, each defining a winding number. This gives a pair of $\mathbb Z$~topological invariants called mirror winding numbers. A detailed calculation of these topological invariants for the Kekulé model can be found in the literature, see for instance~\cite{Kariyado2017} or the appendix of~\cite{Coutant2024}. The result is: 
\begin{subequations}
	\begin{eqnarray}
		(n_S, n_A) &=& (0,0) \quad \text{if} \quad s>t , \\
		(n_S, n_A) &=& (1,-1) \quad \text{if} \quad s<t .
	\end{eqnarray}
\end{subequations}
In the topological phase $(n_S, n_A) = (1,-1)$, the system displays edge waves along symmetry preserving edges~\cite{Kariyado2017,Coutant2024}. In Appendix~\ref{App:SquareLattice}, we compute these topological invariant in a square lattice model. 

As was shown in~\cite{Juricic2012,Kariyado2019} this topology can also be probed by a defect consisting in an isolated hexagon threaded by a $\pi$-flux (see Fig.~\ref{Fig:Kekule_PiF}-(d)). Notice that we consider here only $\pi$-flux threading a ``molecular hexagon'', that is with intramolecular hopping all equal to $s$. This is because the other type of hexagon would break mirror symmetry~\cite{Kariyado2019}. The $\pi$-flux is obtained by flipping the sign of hopping coefficients along a line starting from the defect (shown in dashed green in Fig.~\ref{Fig:Kekule_PiF}-(d)). This does not break mirror symmetry, but the mirror operation now has to be combined with a gauge transformation (see Appendix~\ref{App:SquareLattice}). Similarly, notice all hexagons besides the defect one can be changed to regular ones by a gauge transformation.

\begin{figure}[htp]
	\centering
	\includegraphics[width=\columnwidth]{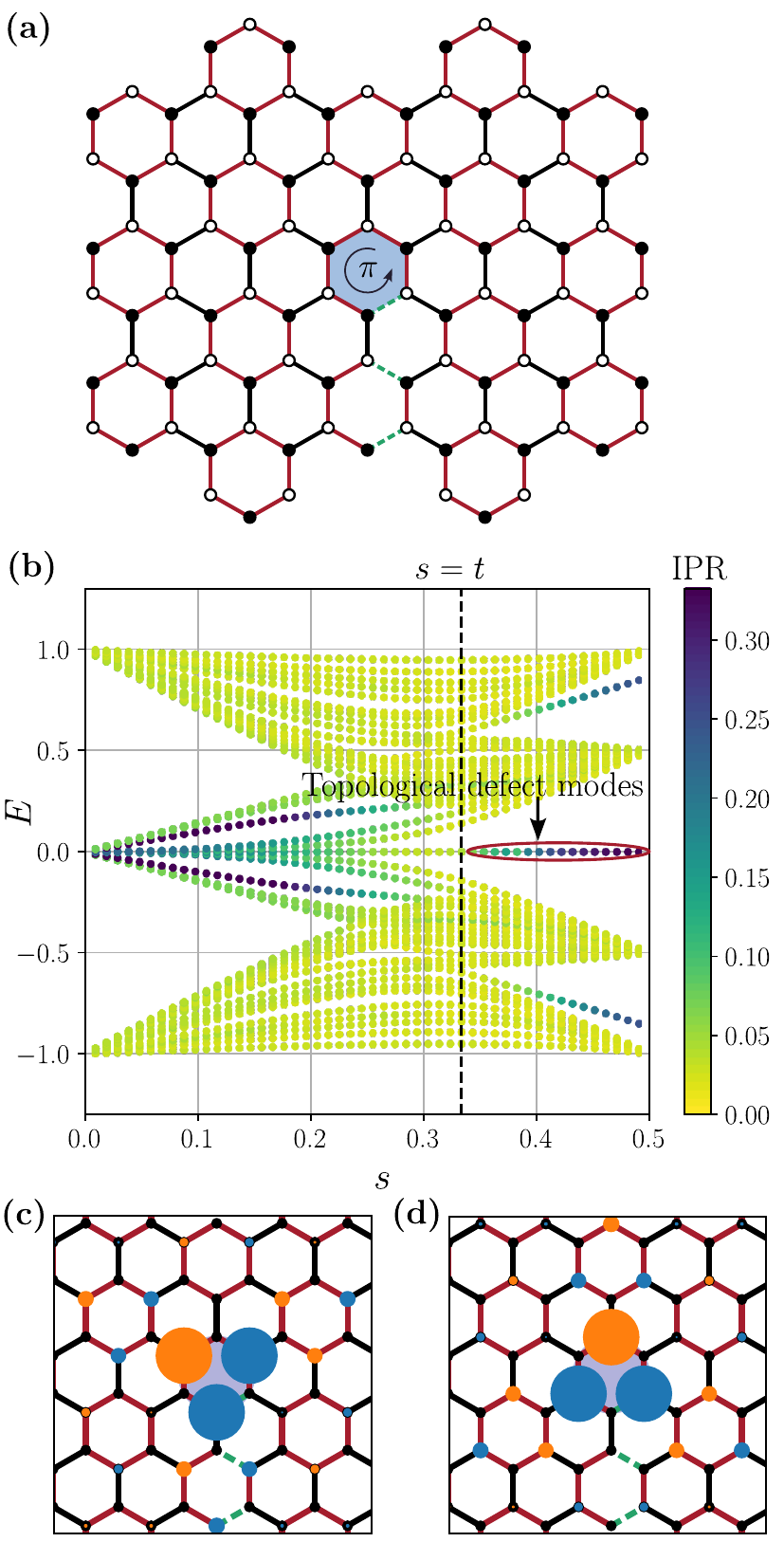} 
	\caption{(a) Finite Kekulé network with a $\pi$-flux in the middle (blue filled hexagon). (b) Spectrum of the network of (a) for varying $s$ and $t = 1 - 2s$. The color scale shows the localization through the inverse participation ratio $\mathrm{IPR} = \sum_j |\psi_j|^4/(\sum_j |\psi_j|^2)^2$. (c-d) Profile of the two zero energy modes of the lattice (a) for $s=0.4$ represented with disks of radius given by the mode amplitude. Blue (resp. orange) disks indicate a positive (resp.negative) amplitude. }
	\label{Fig:PiKekule_TopoTrans} 
\end{figure}

When considering an isolated $\pi$-flux, a pair of localized modes around the defect arise at $E=0$, each with support on one of the two sublattices. Somewhat surprisingly, the pair of topological defect modes is present in the trivial phase of the bulk structure, i.e. when $(n_S, n_A) = (0,0)$. This is illustrated in Fig.~\ref{Fig:PiKekule_TopoTrans}. This can be understood by appealing to a bulk-defect correspondence~\cite{Teo2010,Zhang2023}, which is analogous to the Jackiw-Rossi mechanism that binds a bound state to a vortex~\cite{Jackiw1981,Hou2007,Gao2019}. To establish it, we need to define a topological index for the defect~\cite{Teo2010,Koshino2014,Coutant2024}. This defect index is the pendant of the vortex winding number in the Jackiw-Rossi model. Here, the defect index is obtained in a similar way to the bulk topological indices: one splits the Hamiltonian of the defect alone into symmetric and anti-symmetric sectors, and compute the difference of number of degrees of freedom $N_{\mathcal A}$ on sublattice $\mathcal A$ and $N_{\mathcal B}$ on sublattice $\mathcal B$. This defines the pair of indices: 
\begin{equation} \label{eq:TopoDefect_Indices}
	(\Delta_S, \Delta_A) = (N_{\mathcal A}^S - N_{\mathcal B}^S, N_{\mathcal A}^A - N_{\mathcal B}^A) . 
\end{equation}
A single hexagon threaded by a $\pi$-flux gives a pair of indices $(\Delta_S, \Delta_A) = (1,-1)$ (see Appendix~\ref{App:TopoDefectIndex}). We now understand the topological defect modes in the following way: when the defect indices differ from the bulk indices, we have $|\Delta_j - n_j|$ topologically protected modes on the symmetry sector $j \in \{A, S\}$. Moreover, if $\Delta_j - n_j>0$ (resp. $<0$), the modes have support on sublattice $\mathcal A$ (resp. sublattice $\mathcal B$). In particular, this explains why the topological defect modes appear in the ``trivial phase'', when $(n_S, n_A) = (0,0)$. This turns out to be a practical advantage: edge waves never coexist with the topological defect modes, as visible in Fig.~\ref{Fig:PiKekule_TopoTrans}-(b). This prevents finite size hybridization, and allows us to work with a reasonably low number of unit cells in the acoustic implementation. 

\begin{figure}[htp]
	\centering
	\includegraphics[width=\columnwidth]{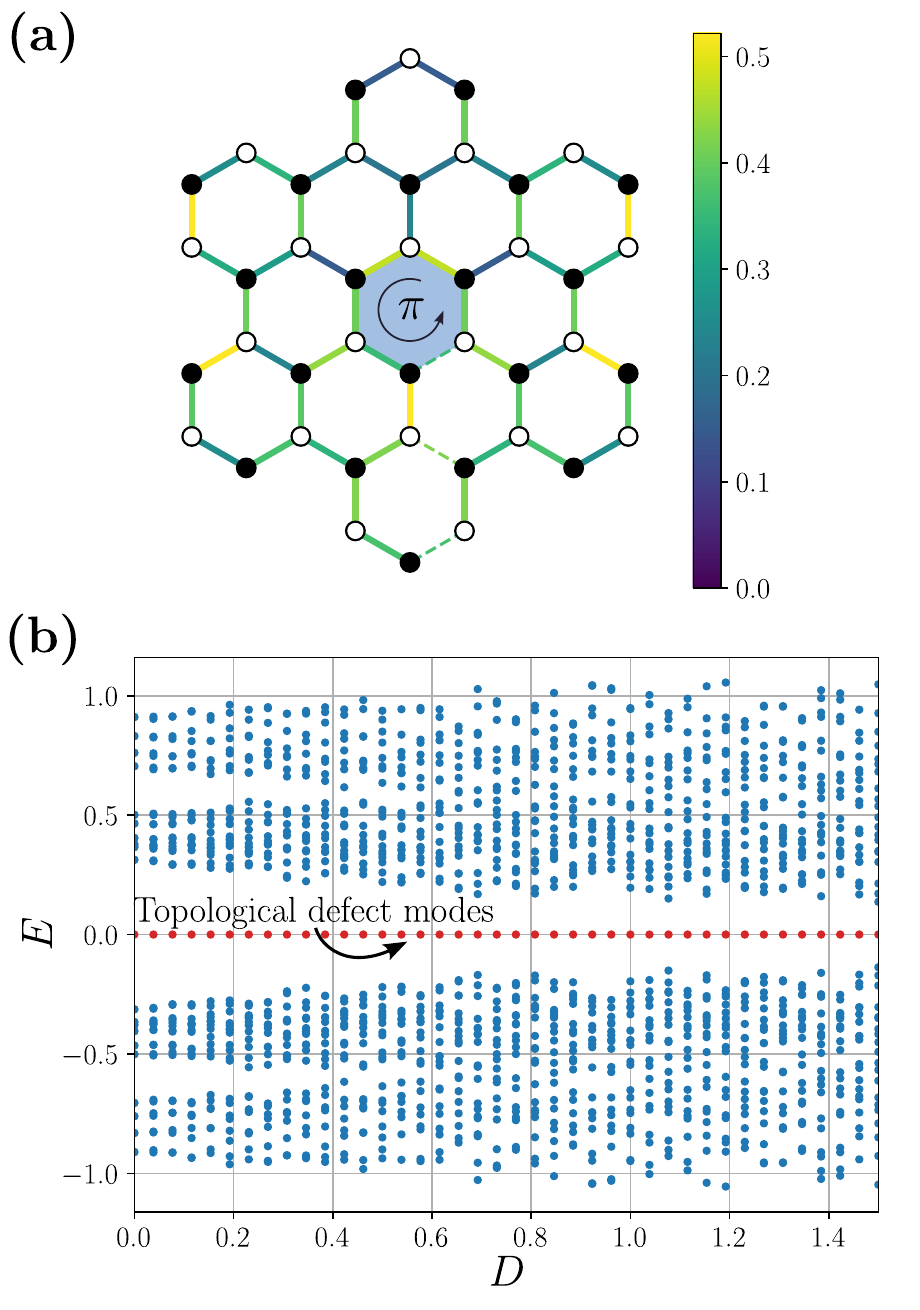} 
	\caption{(a) Finite Kekulé network with a $\pi$-flux in the middle (blue filled hexagon). Colorbar indicate the hopping values for the highest disorder strength $D=1.5$. Negative hoppings are emphasized with dashed lines. (b) Spectrum of the network of (a) for increasing disorder strength. All hopping are taken randomly in the intervals $[s_0, s_0+D]$ for intracell and $[t_0, t_0+D]$ for extracell and symmetrized with respect to the vertical axis afterwards. We took $s_0 = 0.4$ and $t_0 = 0.2$.}
	\label{Fig:PiKekule_Disorder} 
\end{figure}

Thanks to the topological protections, the defect modes are robust to any symmetry preserving perturbation, while their eigenenergy is pinned at $E=0$. We illustrate this by considering a (finite) network with an isolated $\pi$-flux and adding a random perturbation on the values of the hopping coefficients while preserving mirror symmetry. Chiral symmetry is also preserved since no new coupling is added nor any on-site potential. This is shown in Fig.~\ref{Fig:PiKekule_Disorder}. We see that the defect modes stays localized with energy $E=0$ even at high disorder strength.

Once again, we emphasized that the bulk-defect correspondence only requires the presence of commuting chiral and mirror symmetries to hold. Hence, similar topological defect modes can be obtained in different lattice models, and will display the same robustness properties (see Appendix~\ref{App:SquareLattice}).

%
%

\section{From acoustic networks to lattice models and $\pi$-fluxes}
\label{Sec_Ac}

Acoustic networks are made of narrow air channels made of tubes of varying lengths and cross-sections. The typical transverse length $\ell_\perp$ of the tubes is assumed much smaller that its typical length $L$ ($\ell_\perp \ll L$) so that inside each tube the propagation is monomodal~\cite{Depollier1990,Wang2017}. We also assume that all tubes have the same length $L$. Under these two assumptions, acoustic networks provide a simple, versatile and broadband platform to realize lattice models, as was shown in~\cite{Zheng2019,Zheng2020,Coutant2020a,Coutant2021,Coutant2021a,Coutant2024}. In this limit, the acoustic pressure inside the tubes satisfies the one-dimensional Helmholtz equation $p'' + k^2 p$ with $k = \om/c_0$, $\om$ the frequency and $c_0$ the speed of sound. The time dependent pressure is given by $\Re(p e^{-i \om t})$. At the nodes of the network, pressure is continuous and flow rate is conserved~\cite{Depollier1990,Chaigne2016}. These conditions are known as the Kirchhoff laws of quantum graphs~\cite{Kuchment2008,Smilansky2007,Delourme2019,Lawrie2022,Lawrie2024,Delourme2024}. 

After integration of the Helmholtz equation along each tube, the acoustic modes of the network are solutions of the eigenvalue problem $E \psi = H \psi$, with an effective Hamiltonian $H$ as in equation~\eqref{eq:H_def}. The vector $\psi = (p_j)$ contains the (complex) pressure amplitudes on the nodes of the network, the eigenvalue $E$ is analogous to the quantum mechanical energy and is related to the acoustic frequency by: 
\begin{equation}
	E = \cos(kL) . 
\end{equation} 
Moreover, the hopping defining the Hamiltonian are given by ratios of the tubes cross sections. We underline that in acoustic networks, the chiral symmetry is obtained by the graph structure and the requirement of equal tube lengths. This allows for a precise control of the chiral symmetry.

To emulate a Kekulé structure, we construct a honeycomb network with hexagons of tubes of cross-section $\sigma_s$ connected to other hexagons by tubes of cross-section $\sigma_t$. The intra- and inter-hexagon hopping are then simply given by 
\begin{equation} \label{eq:AcKekule_Hop}
	s = \frac{\sigma_s}{\sigma_t + 2\sigma_s}, \qquad \text{and} \qquad t = \frac{\sigma_t}{\sigma_t + 2\sigma_s} . 
\end{equation}
We refer to the literature~\cite{Zheng2019,Zheng2020,Coutant2020a,Coutant2021,Coutant2021a,Coutant2024} for more details on the derivation. 

At this level, the difficulty is to thread a chosen hexagon with a $\pi$-flux. Indeed, although hoppings can be easily tuned through the tubes sections, equation~\eqref{eq:AcKekule_Hop} shows that the hoppings are manifestly positive. To obtain negative hoppings, we consider a double layered network, composed of two identical layers. Now, each tube can either connect the same layer (straight coupling) or different layers (cross coupling). Explicitly, if we call $p_j^{u}$ the pressure amplitude on node $j$ of the upper layer, and $p_j^{l}$ the pressure amplitude on node $j$ of lower layer, the eigenvalue problem has the general form: 
\begin{subequations}
	\begin{eqnarray}
		E p_j^{u} &=& \sum \alpha_{jj'} p_{j'}^{u} + \sum \gamma_{jj'} p_{j'}^{l} , \\
		E p_j^{l} &=& \sum \alpha_{jj'} p_{j'}^{l} + \sum \gamma_{jj'} p_{j'}^{u} , 
	\end{eqnarray}
\end{subequations}
where $\alpha_{ij}$ are the straight couplings and $\gamma_{jj'}$ the cross-coupling. Now, since the two layers are identical, we can look for solutions that are either symmetric, $p_j^S = p_j^{u} = p_j^{l}$, or anti-symmetric, $p_j^A = p_j^{u} = - p_j^{l}$, with respect to the layer exchange symmetry. We obtain two equations of the general form: 
\begin{subequations} \label{eq:Dlay_Hop}
	\begin{eqnarray}
		E p_j^S &=& \sum \alpha_{jj'} p_{j'}^S + \sum \gamma_{jj'} p_{j'}^S , \\
		E p_j^A &=& \sum \alpha_{jj'} p_{j'}^A - \sum \gamma_{jj'} p_{j'}^A . 
	\end{eqnarray}
\end{subequations}
We see that cross-couplings contribute as negative hoppings in the anti-symmetric sector. Therefore, to realize a Kekulé network with a $\pi$-flux defect as in Fig.~\ref{Fig:Kekule_PiF}-(d), we construct a double layer honeycomb acoustic network, with cross-couplings where hoppings must be negative, and straight couplings elsewhere. Moreover, we notice that the (layer exchange) symmetric sector of the same structure reduces to a Kekulé network without the $\pi$-flux defect. Hence, in the trivial phase all symmetric modes lie in the bands, which means in-gap modes can only be anti-symmetric and hence, due to the presence of the $\pi$-flux defect.

\section{Experiment}

\begin{figure}[htp]
	\centering
	\includegraphics[width=\columnwidth]{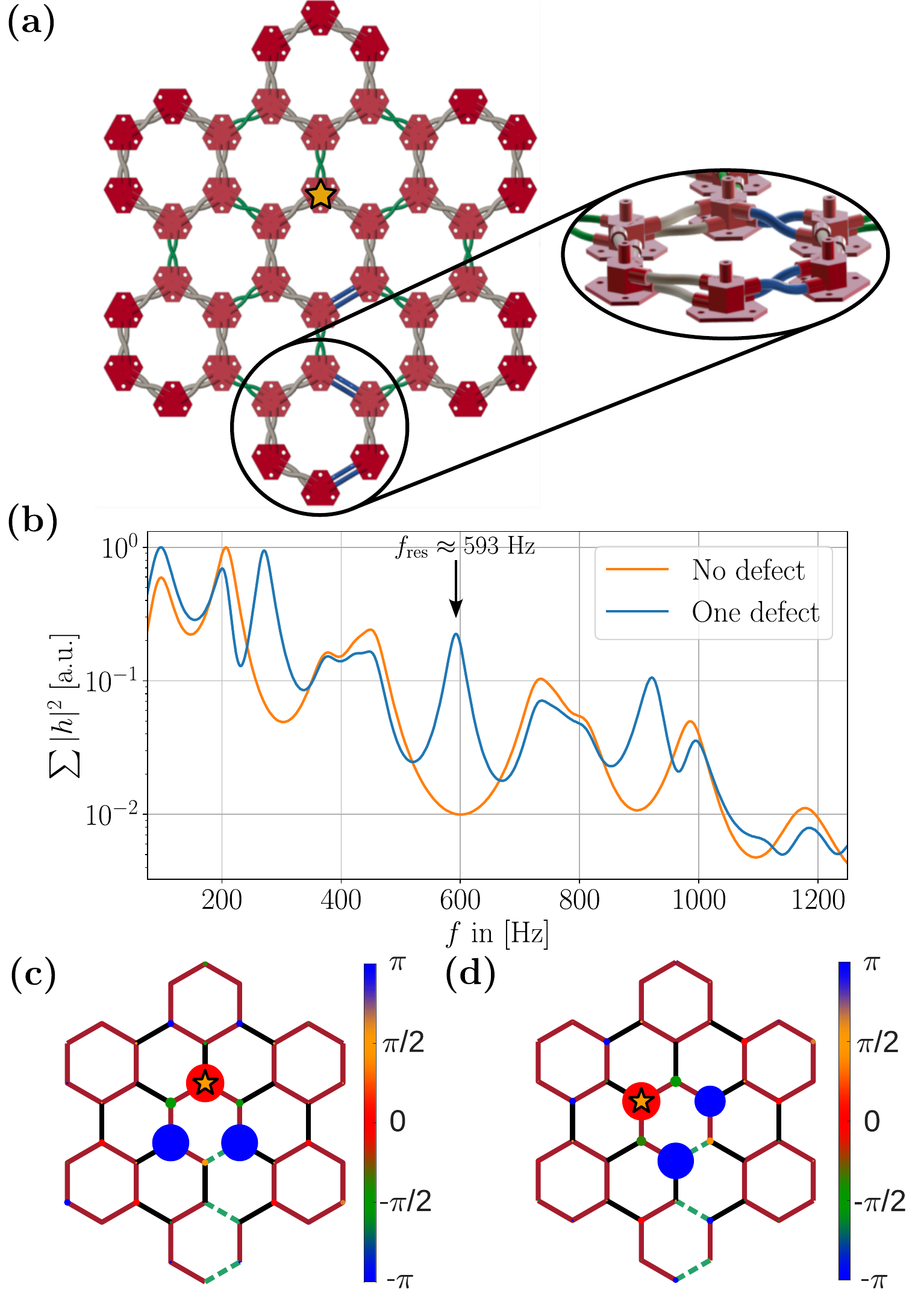} 
	\caption{(a) Schematic of the experimental setup, in the configuration with the $\pi$-flux defect. Connecting modules are in red, gray tubes have section $\sigma_s$ while green tubes have a section $\sigma_t$. Tubes in blue are inter-layer connected, hence leading to negative hopping coefficients (see equation~\eqref{eq:Dlay_Hop}). Inset shows a zoom to see the different structures of intra- and inter-layer coupling tubes. (b) Frequency Response Function (see equation~\eqref{eq:FRF_def}) of a simple cavity (7 hexagons) with (blue) or without (orange) $\pi$-flux in the middle. The frequency of the topological defect mode is estimated at $f_{\rm res} = 593 \pm 10 \; \mathrm{Hz}$, as indicated by the arrow. Source location is shown with a star in (a). (c-d) Pressure response function profile at $f=593$ Hz. The disk radii are proportional to the modulus and the color shows the phase.}
	\label{Fig:Exp_PiKekule} 
\end{figure}

We constructed a double layer network with 7 molecular hexagons, as shown in Fig.~\ref{Fig:Exp_PiKekule}-(a). The network is made with silicon tubes of length $L = 14.4$~cm and inner diameters of 8.2~mm and 4.8~mm. The cross section ratio gives us the hopping coefficients through equation~\eqref{eq:AcKekule_Hop}: $s \approx 0.43$ and $t\approx 0.15$. The tubes are then connected by 3D-printed ABS modules. The tube length $L$ is obtained by adding the length of a tube alone ($10.8$~cm) to twice the length of the tube part inside the module ($1.8$~cm), as explained p.~364 of~\cite{Chaigne2016}. Each module has two meeting points connecting three tubes (two on the edges): one for the top network and one for the bottom network. The top and bottom meeting points are not connected. Each module hosts a microphone that measures the pressure on the upper layer with a sensitivity of 24 mV/Pa. The system is excited by a compression chamber (BMS 4552-8) that can be connected to any module on the top layer. 

We then measure the response of the system to a frequency swipe over the range $50$~Hz to $2000$~Hz. The method of~\cite{Novak2015} is used to extract each pressure frequency response $h_j$ on each top layer node~\footnote{For practical reasons, two sites on the periphery did not host a microphone, hence, no amplitude is shown (see Appendix~\ref{App:MoreModes}). However, these amplitudes are very small.} $j$, and remove effects due to source non-linearities. We then define the power frequency response function of the lattice as 
\begin{equation} \label{eq:FRF_def}
	F = \sum_{j} |h_j|^2 . 
\end{equation}
We perform the experiment twice: one without $\pi$-flux (all tubes connect the same layers) and one with $\pi$-flux (there are cross-connecting tubes). The two frequency response functions are shown in Fig.~\ref{Fig:Exp_PiKekule}-(b). We clearly see the bands and gap of the system without $\pi$-flux defect, in particular in the range around $f_0 = c_0/(4L) = 596 \pm 10 \; \mathrm{Hz}$, which corresponds to $E=0$ (uncertainty for $f_0$ comes from measuring the length $L$). Moreover, the network with a $\pi$-flux defect has a resonance peak near $f_0$, which confirms the presence of the topological modes. The frequency of the topological defect modes is estimated at $f_{\rm res} = 593 \pm 10 \; \mathrm{Hz}$ (uncertainty for $f_{\rm res}$ comes from the broadness of the resonance peak). The agreement is quite remarkable, thereby confirming the presence of the topological defect modes. In Fig.~\ref{Fig:Exp_PiKekule}-(c,d), we show the profiles of the two modes by changing the source position (in appendix~\ref{App:MoreExp} we show the power frequency response function for all source positions). This also compares remarkably with Fig.~\ref{Fig:PiKekule_TopoTrans}.

It is also interesting to notice that additional defect modes appear in the other gaps (near $f = 270$~Hz and $f = 920$~Hz). These modes (shown in Appendix~\ref{App:MoreModes}) are also present in the lattice model, as can be seen in Fig.~\ref{Fig:PiKekule_TopoTrans}-(b), but are not protected by chiral and mirror symmetry. Whether these extra defect modes are protected by a different topological invariant, possibly relying on other spatial symmetries, is an open question that goes beyond the scope of the present work.

%
%
\section{Conclusion and outlook}

In this work we showed how to obtain topological defect modes in a two-dimensional lattice displaying both mirror and chiral symmetry. 
The presence of these modes is explained by an appropriate bulk-defect correspondence relying on the topological indices of the defect (see equation~\ref{eq:TopoDefect_Indices}). Moreover, as illustrated in Fig.~\ref{Fig:PiKekule_Disorder}, these modes have remarkable robustness to symmetry preserving disorder, with their eigenenergy pinned at zero. 

The bulk-defect correspondence is confirmed experimentally in a double layer acoustic network (see Fig.~\ref{Fig:Exp_PiKekule}). Unlike approaches based on the tight binding approximation, the advantage of acoustic networks is that they allow us to realize discrete lattice models over a large frequency range, and with easy tunable hopping coefficients~\cite{Zheng2019,Coutant2021a}. By using a double layer structure, we show that (effective) negative hopping coefficients can be emulated by inter-layer couplings. 

This work provides the foundation for new ways of localizing wave energy in a robust fashion. It could also be used to robustly guide waves along complex paths by using lines of $\pi$-flux defects~\cite{Kariyado2019}.

%
%

\section*{Aknowledgements}

We would like to thank T.~Torres and V.~Achilleos for interesting discussions on the topic. 

\section*{Author contributions}

A.~C. conceived the model and the theoretical analysis. 
R.~C. and M.~P. conceived the experiment, the experimental measurement and data analysis. 
All authors initiated the project, discussed the results and implications, and commented on the manuscript at all stages.

\bibliography{/home/acoutant/Nextcloud_CNRS/Recherche/Notes/SuperBib.bib}

%
%
\appendix


\section{Topological modes on a $\pi$-flux defect in a square lattice}
\label{App:SquareLattice}

\begin{figure}[htp]
	\centering
	\includegraphics[width=\columnwidth]{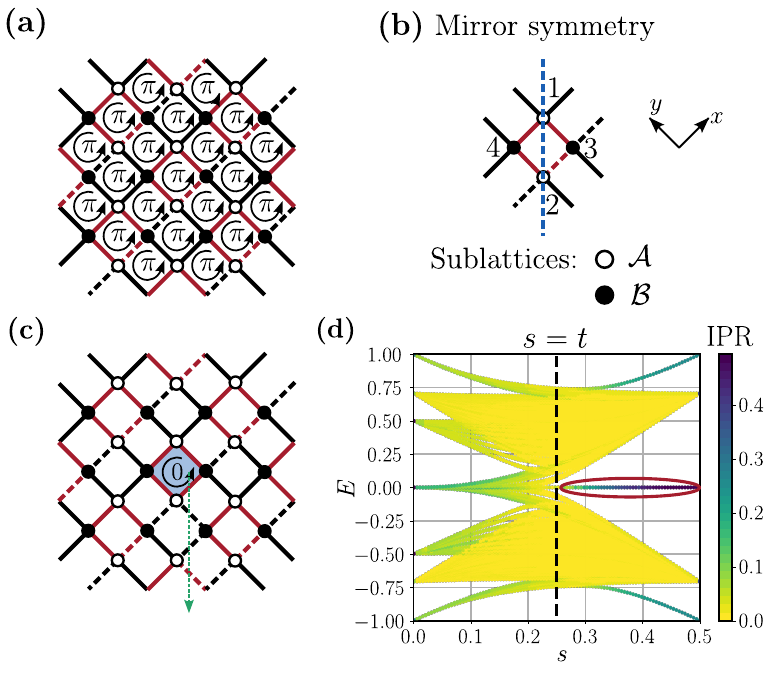} 
	\caption{(a) Illustration of the model on a square lattice. As in Fig.~\ref{Fig:Kekule_PiF}, red links correspond to hoppings with absolute value of $s$ and black ones to $t$. Positive (resp. negative) hoppings are shown with a solid (resp. dashed) line. (b) Unit cell of the model. The mirror symmetry axis is emphasized with a blue dashed line. (c) Lattice with an isolated defect made of a plaquette with zero flux (emphasized in light blue). (d) Spectrum of a finite lattice with 100 square molecules for varying $s$ and $t = 1/2-s$.}
	\label{Fig:BBH_PiF} 
\end{figure}

The same type of topological defect modes can be obtained in a square lattice. For this, we consider the tight binding model depicted in Fig.~\ref{Fig:BBH_PiF}-(a). This model has been consider in many works as an example of higher order topological insulator~\cite{Benalcazar2017a,SerraGarcia2018,Mittal2019,Qi2020}. In an appropriate basis (see Fig.~\ref{Fig:BBH_PiF}-(b)), it is described by the Bloch Hamiltonian: 	
\begin{equation} \label{eq:BBH_BHam}
	H(q_x,q_y) = 
	\bmat 0 & Q \\
	Q^\dagger & 0 \\
	\emat 
\end{equation}
with the block matrix 
\begin{equation}
	Q = \bmat s+t e^{i q_y} & s + t e^{i q_x} \\
	-s-t e^{-i q_x} & s + t e^{-i q_y} \\
	\emat 
\end{equation}
and where $q_x$ and $q_y$ are the Bloch wavevector components (see Fig.~\ref{Fig:BBH_PiF}-(b)). Because of the $\pi$-fluxes, the mirror symmetry operation $M$ is a combination of a geometric and a gauge transformation: 
\begin{equation} \label{eq:Mirror_BBH}
	M = M_g G , 
\end{equation}
with 
\begin{equation} \label{eq:BBH_Mirror_Mat}
	M_g = \bmat 1 & 0 & 0 & 0 \\ 0 & 1 & 0 & 0 \\ 0 & 0 & 0 & 1 \\ 0 & 0 & 1 & 0 \emat \quad \text{and} \quad G = \mathrm{diag}(1,-1,1,1). 
\end{equation}

We now show how to obtain the mirror winding numbers, as was done for the Kekulé model in~\cite{Kariyado2017,Coutant2024}. For this we first consider the mirror invariant line of the first Brillouin zone: $q_x = q_y = q$. Along this line, the Hamiltonian~\eqref{eq:BBH_BHam} commutes with the mirror symmetry, i.e. $[H(q_x,q_x), M] = 0$. We can then project the Bloch Hamiltonian on the symmetric and anti-symmetric sectors. Explicitly, the symmetric subspace is spanned by 
\begin{equation} \label{eq:BBH_basis_Sym}
	\bmat 1 \\ 0 \\ 0 \\ 0 \emat \quad \text{and} \quad \frac1{\sqrt 2} \bmat 0 \\ 0 \\ 1 \\ 1 \emat , 
\end{equation}
while the anti-symmetric subspace is spanned by
\begin{equation} \label{eq:BBH_basis_ASym}
	\bmat 0 \\ 1 \\ 0 \\ 0 \emat \quad \text{and} \quad \frac1{\sqrt 2} \bmat 0 \\ 0 \\ -1 \\ 1 \emat . 
\end{equation}
Notice that $\bmat 0 & 1 & 0 & 0 \emat^T$ is in the anti-symmetric subspace because of the gauge part of the mirror operation~\eqref{eq:Mirror_BBH}.In other words, the $\pi$-flux changes the dimensionalities of the symmetric and anti-symmetric subspaces. Using the four vectors in \eqref{eq:BBH_basis_Sym}, \eqref{eq:BBH_basis_ASym} as a basis, the Bloch Hamiltonian \eqref{eq:BBH_BHam} has a block form
\begin{equation}
	\bmat H_S(q) & 0 \\ 0 & H_A(q) \emat , 
\end{equation}
where $H_S(q)$ and $H_A(q)$ are the projected Hamiltonians on the symmetric and anti-symmetric subspaces. A direct calculation leads to: 
\begin{subequations}
	\begin{eqnarray}
		H_S(q) &=& \frac1{\sqrt 2} \bmat 0 & s + t e^{iq} \\ s + t e^{-iq} & 0 \emat , \\
		H_A(q) &=& \frac1{\sqrt 2} \bmat 0 & s + t e^{-iq} \\ s + t e^{iq} & 0 \emat . 
	\end{eqnarray}
\end{subequations}
We see that we obtain two copies of the Su-Schrieffer-Heeger (SSH) Hamiltonian (up to the $1/\sqrt 2$ overall factor), but the off-diagonal element rotates in opposite sense in the complex plane. This guarantees that the two winding numbers are opposite. Explicitly, as in the case of the SSH model~\cite{Asboth2016}, we obtain: 
\begin{subequations}
	\begin{eqnarray}
		(n_S, n_A) &=& (0,0) \quad \text{if} \quad s>t , \\
		(n_S, n_A) &=& (1,-1) \quad \text{if} \quad s<t . 
	\end{eqnarray}
\end{subequations}
Now, since in this model, all plaquettes are threaded by a $\pi$-flux, a defect is made of an isolated plaquette with no flux, as shown in Fig.~\ref{Fig:BBH_PiF}-(c). It is obtained similarly to the case of the Kekulé model, that is, by flipping the signs of hopping coefficients along a semi-infinite line (see Fig.~\ref{Fig:BBH_PiF}-(c)). 

We now investigate the presence of topological defect modes. For this, we consider a large finite lattice for varying hopping coefficients (without loss of generality, we impose the relation $2s+2t = 1$, as is the case in an acoustic network). We observe the same mechanism as in the Kekulé model in the core of the text (e.g. Fig.~\ref{Fig:PiKekule_TopoTrans}): in the trivial phase $(n_S, n_A) = (0,0)$ there is a pair of zero energy modes localized on the defect, as we see in Fig.~\ref{Fig:BBH_PiF}-(d). In the topological phase on the contrary, no defect mode is present. Due to the finite size of the lattice, we observe edge and corner modes at the outer edges of the system (see Fig.~\ref{Fig:BBH_PiF}-(d)). As in the main text, this is well understood from the bulk-defect correspondence, since an isolated $0$-flux plaquette in this model has a topological index of: 
\begin{equation}
	(\Delta_S, \Delta_A) = (1,-1). 
\end{equation}

\section{Topological defect indices}
\label{App:TopoDefectIndex}

We now provide more details on how to compute these topological defect in both the Kekulé and square lattice model. We start with the latter, since the derivation of the defect indices is very similar to that of the topological invariants, as presented in Appendix~\ref{App:SquareLattice}. 

\begin{figure}[htp]
	\centering
	\includegraphics[width=\columnwidth]{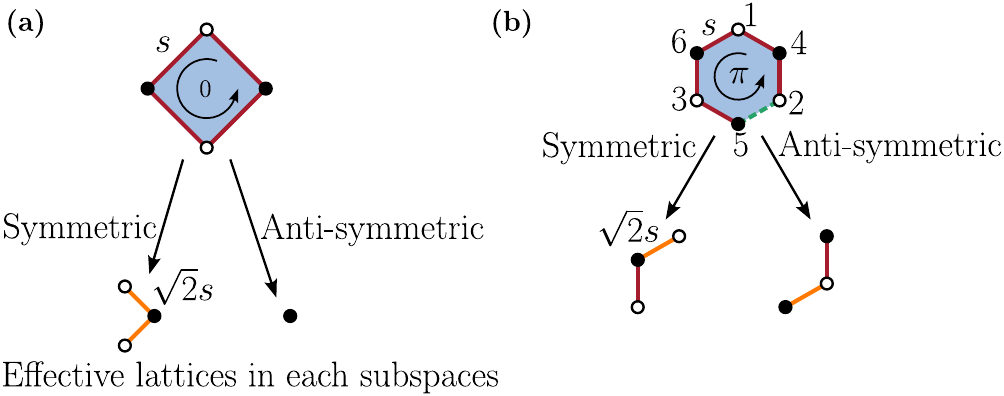} 
	\caption{Illustration of the symmetry decomposition of an (isolated) defect. (a) $0$-flux plaquette in the square lattice model. (b) $\pi$-flux plaquette in Kekulé model. }
	\label{Fig:TopoIndex_Defects} 
\end{figure}

To proceed, we consider an isolated defect, as shown in Fig.~\ref{Fig:TopoIndex_Defects}. In the square lattice case, it consists in a square molecule threaded with no flux (see Fig.~\ref{Fig:TopoIndex_Defects}-(a)). The Hamiltonian of such an isolated defect reads: 
\begin{equation}
	H_{\rm defect} = s \bmat 0 & 0 & 1 & 1 \\
	0 & 0 & 1 & 1 \\
	1 & 1 & 0 & 0 \\
	1 & 1 & 0 & 0\\
	\emat
\end{equation}
Now, because of the absence of $\pi$-flux, the mirror symmetry has no gauge part this time, hence it is just given by $M_g$ in \eqref{eq:BBH_Mirror_Mat}. Now, similarly to Appendix~\ref{App:SquareLattice}, we split the amplitude vector space into a symmetric and an anti-symmetric subspace. Using the expression for $M_g$, the symmetric subspace is spanned by 
\begin{equation} 
	\bmat 1 \\ 0 \\ 0 \\ 0 \emat, \bmat 0 \\ 1 \\ 0 \\ 0 \emat \quad \text{and} \quad \frac1{\sqrt 2} \bmat 0 \\ 0 \\ 1 \\ 1 \emat , 
\end{equation}
while the anti-symmetric one is spanned by 
\begin{equation}
	\frac1{\sqrt 2} \bmat 0 \\ 0 \\ 1 \\ 1 \emat . 
\end{equation}
We now obtain the projected Hamiltonians onto those subspaces: 
\begin{subequations}
	\begin{eqnarray}
		H_{\rm defect}^S &=& \sqrt 2 s \bmat 0 & 1 & 0\\ 1 & 0 & 1 \\ 0 & 1 & 0 \emat , \\
		H_{\rm defect}^A &=& s . 
	\end{eqnarray}
\end{subequations}
These Hamiltonians can be seen as tight binding Hamiltonians of two effective lattices, as shown in Fig.~\ref{Fig:TopoIndex_Defects}-(a). Since the mirror symmetry commute with the chiral operator $\Gamma$, these two lattices have the sublattice symmetry. Now, the indices are given by the difference in sublattice sites, i.e. equation~\eqref{eq:TopoDefect_Indices}. A more algebraic way to phrase this is to consider $\Gamma_j$ with $j \in \{S, A\}$, the matrices induced on the symmetric and anti-symmetric subspaces (which is well-defined because $\Gamma$ commutes with $M_g$). Then the indices are given by~\cite{Koshino2014}: 
\begin{equation}
	\Delta_j = \mathrm{Tr}(\Gamma_j) . 
\end{equation}
In the case of the square molecule, we can directly see from Fig.~\ref{Fig:TopoIndex_Defects}-(a) that: 
\begin{equation}
	(\Delta_S, \Delta_A) = (1,-1) , 
\end{equation}
as mentioned in Appendix~\ref{App:SquareLattice}. 

The same procedure can be done for an isolated defect in the Kekulé model. This time, the defect is a hexagonal molecule threaded by a $\pi$-flux, as illustrated in Fig.~\ref{Fig:BBH_PiF}-(b). The Hamiltonian is given by: 
\begin{equation}
	H_{\rm defect} = s \bmat 0 & 0 & 0 & 1 & 0 & 1 \\
	0 & 0 & 0 & 1 & -1 & 0 \\
	0 & 0 & 0 & 0 & 1 & 1 \\
	1 & 1 & 0 & 0 & 0 & 0 \\
	0 & -1 & 1 & 0 & 0 & 0 \\
	1 & 0 & 1 & 0 & 0 & 0 \\
	\emat 
\end{equation}
Notice that we added the site numbering in Fig.~\ref{Fig:BBH_PiF}-(b). Now, the mirror symmetry is given by the operation:
\begin{equation}
	M = \bmat 1 & 0 & 0 & 0 & 0 & 0 \\ 
	0 & 0 & 1 & 0 & 0 & 0 \\ 
	0 & 1 & 0 & 0 & 0 & 0 \\ 
	0 & 0 & 0 & 0 & 0 & 1 \\ 
	0 & 0 & 0 & 0 & -1 & 0 \\ 
	0 & 0 & 0 & 1 & 0 & 0 \\ 
	\emat . 
\end{equation}
Again, this is a combination of a geometric and a gauge transformation. We can now split the space into a symmetric subspace, spanned by 
\begin{equation} 
	\bmat 1 \\ 0 \\ 0 \\ 0 \\ 0 \\ 0 \emat , \frac1{\sqrt{2}} \bmat 0 \\ 1 \\ 1 \\ 0 \\ 0 \\ 0 \emat \quad \text{and} \quad \frac1{\sqrt{2}} \bmat 0 \\ 0 \\ 0 \\ 1 \\ 0 \\ 1 \emat 
\end{equation}
and an anti-symmetric subspace, spanned by 
\begin{equation} 
	\bmat 0 \\ 0 \\ 0 \\ 0 \\ 1 \\ 0 \emat , \frac1{\sqrt{2}} \bmat 0 \\ 1 \\ -1 \\ 0 \\ 0 \\ 0 \emat \quad \text{and} \quad \frac1{\sqrt{2}} \bmat 0 \\ 0 \\ 0 \\ 1 \\ 0 \\ -1 \emat 
\end{equation}
Notice once again the effect of the $\pi$-flux: the vector $\bmat 0 & 0 & 0 & 0 & 1 & 0 \emat^T$ is the anti-symmetric subspace. After projection, we obtain two Hamiltonians: 
\begin{subequations}
	\begin{eqnarray}
		H_{\rm defect}^S &=& s \bmat 0 & 0 & \sqrt 2 \\ 0 & 0 & 1 \\ \sqrt 2 & 1 & 0 \emat , \\
		H_{\rm defect}^A &=& s \bmat 0 & 1 & \sqrt 2 \\ 1 & 0 & 0 \\ \sqrt 2 & 0 & 0 \emat . 
	\end{eqnarray}
\end{subequations}
Their corresponding lattices are shown in Fig.~\ref{Fig:TopoIndex_Defects}-(b). We can now directly read the two topological indices: 
\begin{equation}
	(\Delta_S, \Delta_A) = (1,-1) , 
\end{equation}
as used in the core of the text.

\section{Defect modes at nonzero energy}
\label{App:MoreModes}

As can be seen in Fig.~\ref{Fig:PiKekule_TopoTrans}, the $\pi$-flux defect in a Kekulé lattice can also host localized modes at nonzero energy. A pair of defect modes has a negative energy, and another has positive energy. Within each pair, there is a symmetric and an anti-symmetric mode with respect to the mirror symmetry. Moreover, the two pairs have opposite energies, and are related by the chiral operator $\Gamma$.

\begin{figure}[htp]
	\centering
	\includegraphics[width=\columnwidth]{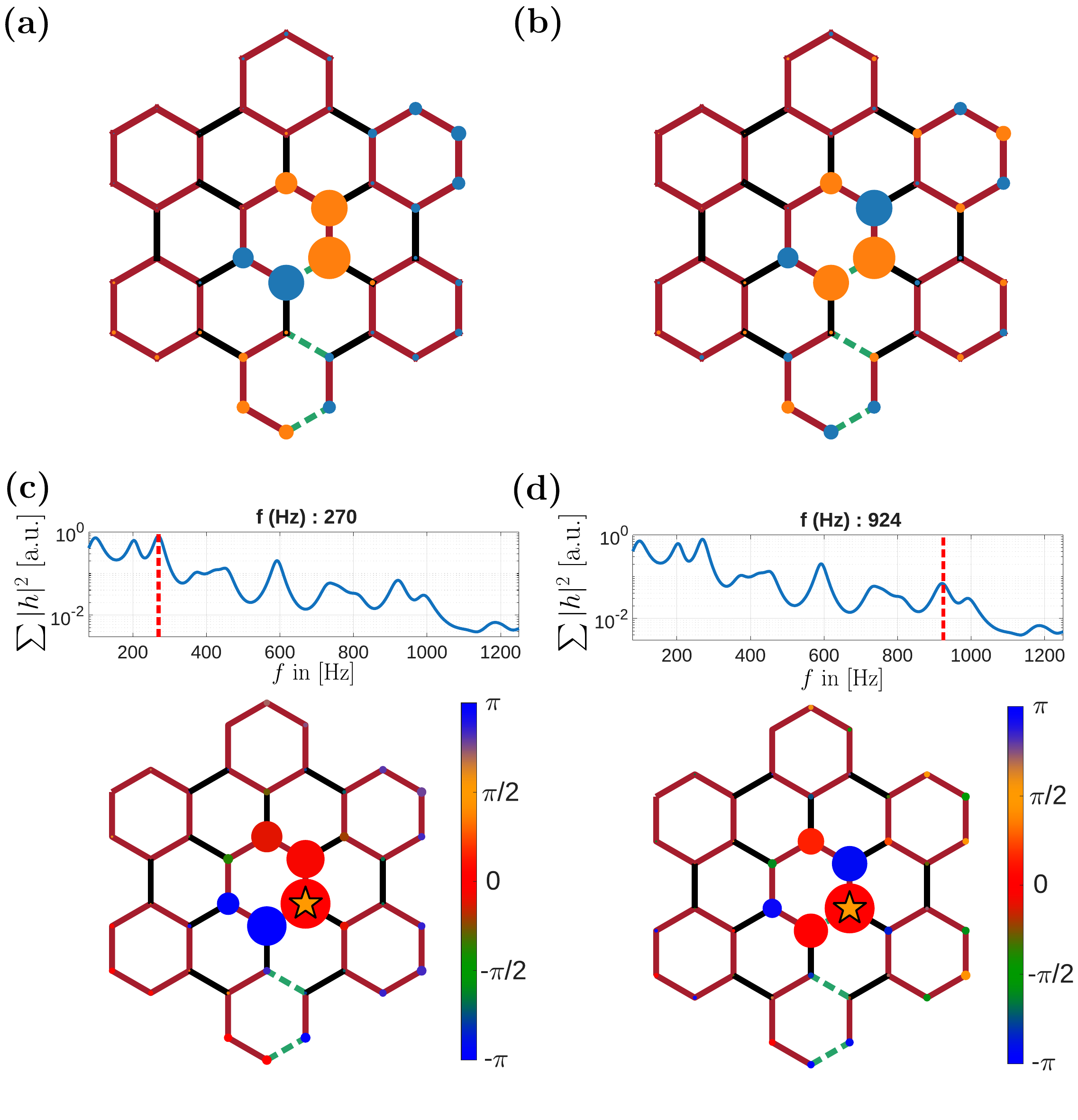} 
	\caption{(a) Defect mode at positive energy. (b) Defect mode at negative energy. (a,b) We took $s=0.43$ and $t=0.14$ as evaluated for the experimental setup. (c) Pressure response function profile at the lower frequency ($E>0$) displaying a defect mode. (d) Pressure response function profile at the lower frequency ($E>0$) displaying a defect mode. (c,d) The source location is shown with a star. The upper inset shows the Frequency Response Function (same as Fig.~\ref{Fig:Exp_PiKekule}-(b)), with the frequency of the source emphasized with a red dashed line.}
	\label{Fig:DefectModes} 
\end{figure}

In Fig.~\ref{Fig:DefectModes}, we compare the modes of the discrete model (a,b) with the ones obtained experimentally (c,d). 

\begin{figure}[htp]
	\centering
	\includegraphics[width=0.6\columnwidth]{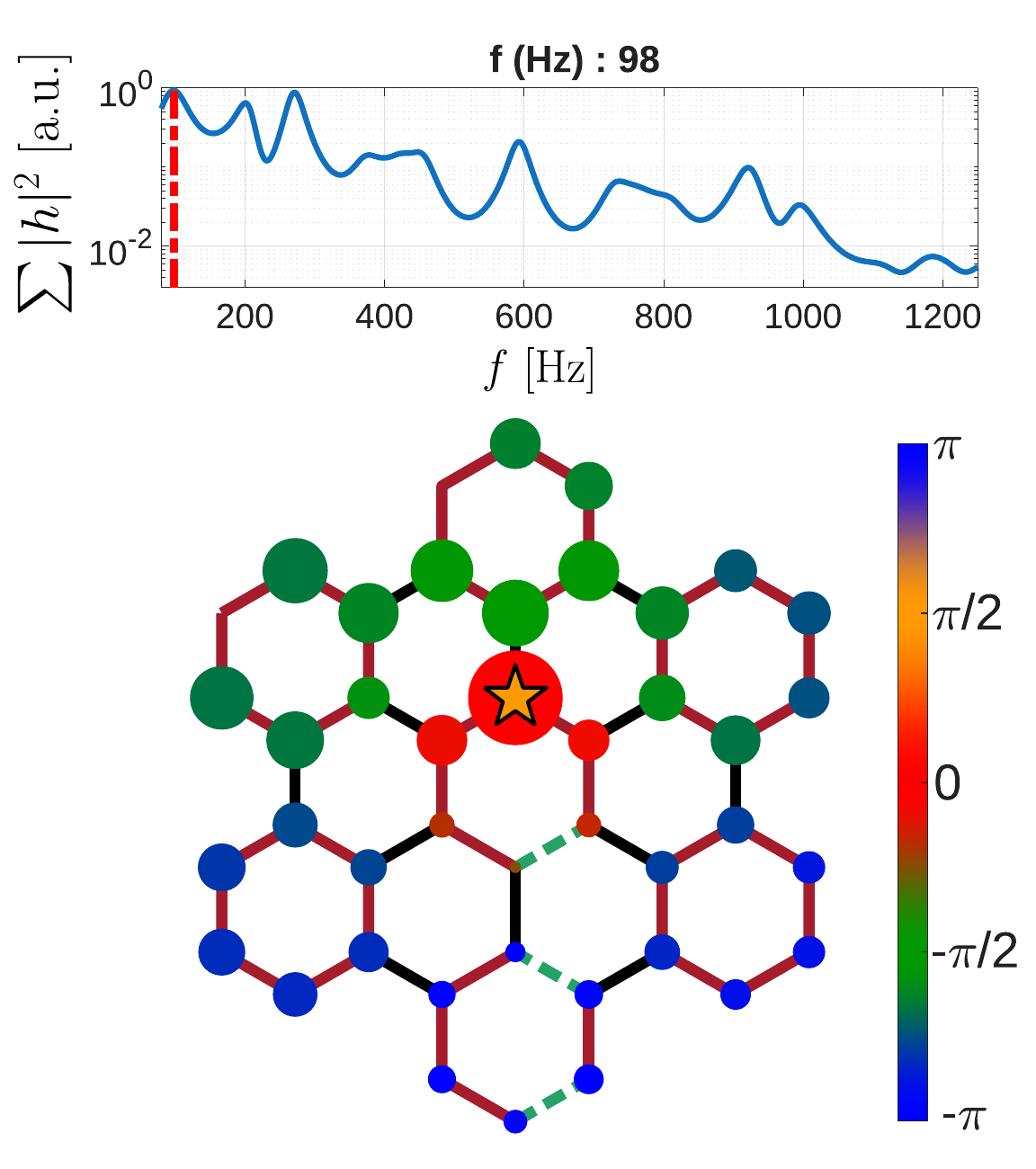} 
	\caption{Pressure response function profile at a low frequency inside a band. The upper inset shows the Frequency Response Function (same as Fig.~\ref{Fig:Exp_PiKekule}-(b)), with the frequency of the source emphasized with a red dashed line. As mentioned in the text, two sites on the periphery did not host a microphone, as is visible here.}
	\label{Fig:BulkMode} 
\end{figure}

To complement these results, in Fig.~\ref{Fig:BulkMode}, we show the profile of the pressure response function at a positive in-band energy. This shows that modes inside a band are delocalized over the system, in contrast to defect localized modes. We notice that the signal decreases on sites further from the source, as expected due to dissipative processes in the system.

\section{Additional informations on the experimental setup}
\label{App:MoreExp}

\begin{figure}[htp]
	\centering
	\includegraphics[width=\columnwidth]{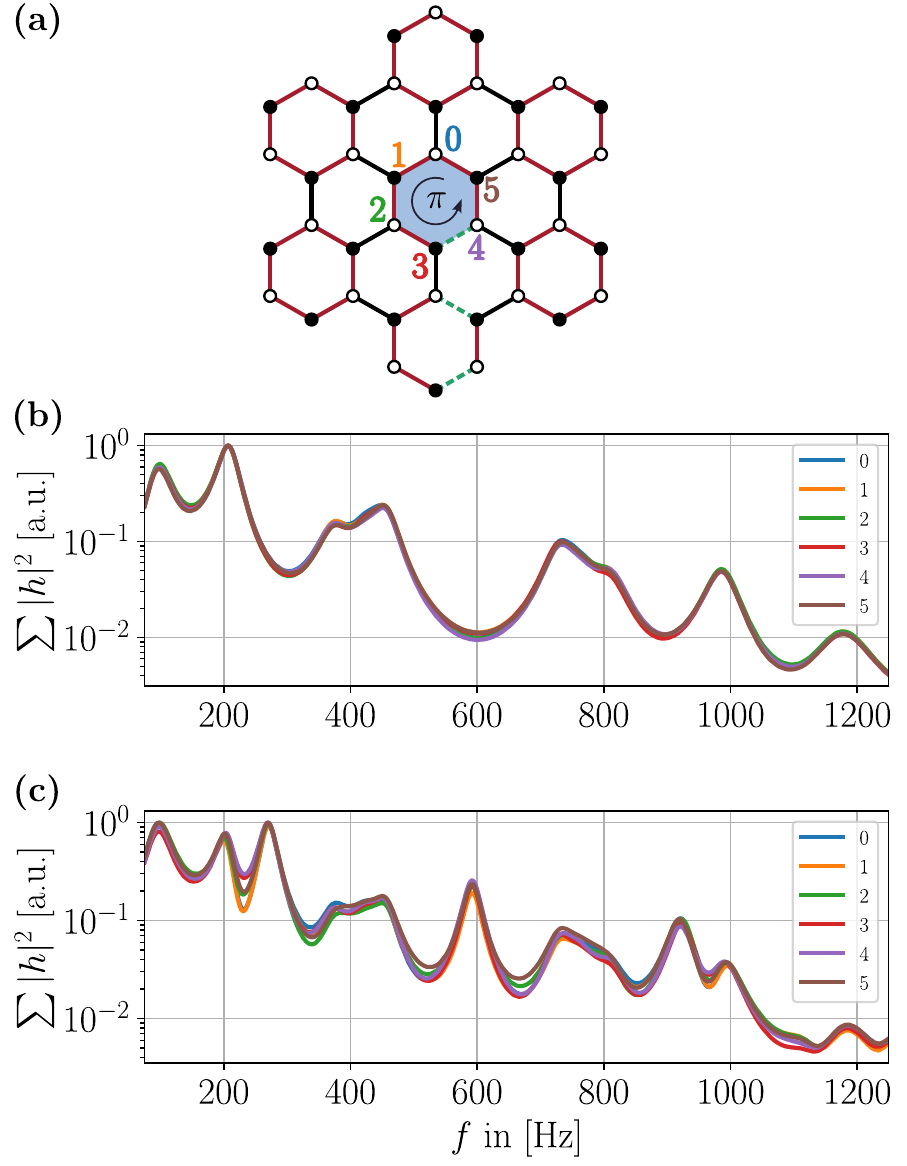} 
	\caption{(a) Schematics of the experimental configuration, with the 6 source positions numbered. (b) Frequency response function for the 6 different source position in a lattice without $\pi$-flux. (c) Frequency response function for the 6 different source positions (legend color matches that of (a)) in a lattice with $\pi$-flux.}
	\label{Fig:Exp_AllSources} 
\end{figure}

The ABS modules are made so as to ensure that the acoustic lengths are identical between modules, regardless of whether the connection is made horizontally or vertically. This is why the tubes are S-shaped: they retain this shape when placed either vertically or horizontally. The various connections are tight-fitting, dismountable, and airtight when vacuum grease is used. This makes it easy to change the positions of the source, microphones, and tube connections.

The microphones and conditioners were co-developed with ALMACOUSTIC~\cite{ALMA} and have a typical sensitivity of 24 mV/Pa. The source output radius is reduced from 1" to 7 mm to fit the vertical wells.

The measurement system, which drives the source and records the microphones, uses a computer and audio-specific devices, including an {\it RME HDSPe MADI} sound card (PCI Express), an {\it RME ADI 648} MADI/ADAT converter, and three AD/DA converters {\it Ferrofish A16 MKII}. The control software was developed in LMA and uses the synchronized recording routine {\it rwAudio} developed by C. Lambourg (ARTEAC-LAB~\cite{ARTEAC-LAB})

To test the symmetry of our system, we run the same experiment as in Fig.~\ref{Fig:Exp_PiKekule} with different source position within the central hexagon. All results are shown in Fig.~\ref{Fig:Exp_AllSources}. When the two layers are uncoupled (no $\pi$-flux), we confirm the hexagonal symmetry to a high accuracy, as all frequency response functions match. When the two layers are coupled (with $\pi$-flux), the imperfection of the symmetry is more visible but are still quite low. This small discrepancy is essentially due to dissipation. Roughly speaking, depending on the source position, a signal has to go down to the bottom layer before going back up, and it loses more energy during that travel compared to staying on the same layer.

We also see that the same resonance peak is found for all position source in the presence of a $\pi$-flux defect. This confirms that the topological defect modes come in pair with support on each sublattice.

\end{document}